\documentclass[aps, prl, reprint]{revtex4-1}

\usepackage{amsmath}
\usepackage[title]{appendix}
\usepackage{graphicx}
\usepackage[section]{placeins}
\usepackage{longtable}

\begin{document}

\title{Why phonon scattering in glasses is universally small at low temperatures} 
\author{Herve M. Carruzzo}
\author{Clare C. Yu}
\affiliation{University of California, Irvine, Irvine, CA 92697}
\date{\today}
\begin{abstract}
We present a novel view of the standard model of tunneling two level systems 
(TLS) to explain the puzzling universal value of a quantity,
$C\sim 3\times 10^{-4}$, that characterizes phonon scattering in glasses
below 1 K as reflected in thermal conductivity, ultrasonic attenuation,
internal friction, and the change in sound velocity. Physical considerations 
lead to a broad distribution of phonon-TLS couplings that 
(1) exponentially renormalize tunneling matrix elements, 
and (2) reduce the TLS density of states through TLS-TLS interactions. 
We find good agreement between theory and experiment for a variety 
of individual glasses.
\end{abstract}

\maketitle

Amorphous solids are ubiquitous and technologically important, yet they still hold mysteries such as the universal values of phonon scattering. Below 1 K, phonon scattering reflected in the thermal conductivity (scaled with natural units) \cite{Freeman_1986},  the internal friction (in the relaxation regime) \cite{Topp_1996}, the change in the sound velocity, and the resonant ultrasonic attenuation \cite{Berret_1988} are quantitatively very similar, regardless of the insulating glassy material. This universality is quite surprising, and, though it has been known for quite some time, remains a puzzle. Why does phonon scattering in these materials show such a lack of sensitivity to their composition and structure?

The standard model of tunneling two level systems (TLS) \cite{Anderson_1972,Phillips_1972}  qualitatively describes the behavior of glasses below 1K.  It postulates the existence of  \emph{independent} entities that tunnel between the two minima of a  double well potential with a wide distribution of tunneling matrix elements  and energy asymmetries. However, this model does not quantitatively explain  the measurements cited above that depend on the coupling of phonons to  tunneling two level systems (TLS). In particular, these measurements all find a rather universal value for a dimensionless coupling  constant, C, given by
\begin{equation}\label{eq:C_TLS}
  C={{\bar P}\gamma^2\over\rho v^2}
  \end{equation}
where ${\bar P}$ is the density of states of tunneling entities,  $\gamma$ is the strength of their coupling with phonons,  $\rho$ is the mass density of the material and  $v$ is the sound velocity given by $v^{-3}={1\over 3}\sum_sv_s^{-3}$,  where $v_s$ is the sound velocity for polarization $s$.  Within the TLS model, the internal friction $Q^{-1}$ is given by $Q^{-1}={\pi\over 2}C$, while the change in sound velocity in the relaxation regime is $\Delta v/v=-\frac{1}{2}C\ln(T/T_o)$ where $T_o$ is an arbitrary reference temperature. Finally, the scaled thermal conductivity \cite{Freeman_1986} is universal because it depends on ratio of the mean free path $\ell$ of a phonon to its wavelength $\lambda$ in the following way:  $\ell/\lambda=1/(2\pi^2 C)$.  Measurements of these quantities find values of C between $2\times 10^{-4}$ and $5\times 10^{-4}$. The universal value of $C$ is quite surprising given that the parameters entering $C$ are nominally independent and vary significantly  from glass to glass. A universal value for this quantity thus implies a  degree of coincidence that strains credulity, as noted by Leggett \cite{Leggett_1991}. Another surprise is the energy scale of the coupling between the sound waves and the TLS which is about 1 eV in insulating glasses, an energy scale that does not match any other in the problem.  

Yu and Leggett  \cite{Yu_1988} (YL) made the first attempt to understand this coincidence. They assumed that phonon mediated interactions between  TLS dominate the physics. While their mean field scenario did indeed explain  the universality, the predicted universal value for $C$ is of order 1,  whereas the observed value is $10^{-4}$.  This failure of a simple mean field theory approach is rather surprising.   What determines the value for $C$?  A variety of rather complicated approaches have been proposed to resolve this question  \cite{Burin_1996,Lubchenko_2008,Parshin_2007,Vural_2011}. These include renormalization group approaches \cite{Burin_1996,Vural_2011}, a random first order phase transition associated with the glass transition \cite{Lubchenko_2008}, two different types of TLS that couple differently to phonons \cite{Schechter_2013}, and vibrational instability of harmonic oscillators associated with the boson peak \cite{Parshin_2007}. These models have been able to arrive at the right order of magnitude for $C$, but the use of a variety of assumptions and estimates have precluded the ability to predict the value of $C$ for different specific glasses.  

We propose a novel explanation based on three aspects implicit in the standard TLS model that were ignored in the original model \cite{Anderson_1972,Phillips_1972} and were only partially considered subsequently. First, the coupling between phonons and TLS implies that the TLS can interact with each other \cite{Yu_1988}. Second, this coupling produces an exponential renormalization of the tunneling matrix element due to phonon overlap between the two wells (a kind of polaron effect) \cite{Kassner_1989}. Third, phonons actually couple to the {\it difference} between the elastic dipole moments in the two wells. If the elastic dipole moment in each well has a random orientation, the difference will also be random and will vary from TLS to TLS, leading to a broad distribution of couplings $\gamma$. Our model explains the universal value of $C$ as well as the observed ($\sim 1$ eV) value of the TLS-phonon coupling at low frequencies. 

We begin by introducing the model for a set of entities that can tunnel between two states, e.g., ``right'' and ``left'' well, randomly distributed in an elastic medium and interacting with phonons:

\begin{equation}\label{eq:H_TLS}
    H=H_{\text{ph}}+{1\over 2}\sum_i 
(\varepsilon_i\sigma^z_i + \Delta_i^o\sigma_i^x) + 
{1\over\sqrt{V}}\sum_{ik}\gamma_i \epsilon^i_{k}\sigma^z_i
\end{equation}
where the free phonon Hamiltonian  $H_{\text{ph}}=\sum_{ks}\hbar\omega(ks)\psi^{\dagger}_{ks}\psi_{ks}$.  $\psi^{\dagger}_{ks}$ and $\psi_{ks}$ are the phonon creation and anihilation operators for wavevector $k$ and polarization $s$,  and $\omega(ks)$ is the phonon dispersion relation \footnote{At high frequencies, the normal modes are not going to be plane waves given the topological disorder. The hope is that this approximation is reasonable for the purpose of calculations in this paper.}. $\varepsilon_i$ is the energy asymmetry between the two wells and $\Delta^o_i$ is the tunneling matrix element of the $i^{\text th}$ TLS. In the TLS-phonon interaction (last term in Eq.(\ref{eq:H_TLS})), $\epsilon^i_{k} = \sum_s\xi_{ks}(i)\psi_{ks}+\xi_{ks}^*(i)\psi_{ks}^{\dagger}$ represents the scalar strain field, where $\xi_{ks}(i)=i\sqrt{ {\hbar\omega(ks)\over 2\rho v^2} } (\sum_{ab}D^{i}_{ab}e^s_{ab}(k))e^{ikr_i}$ and $\gamma_i D_{ab}^{i}$ is the TLS elastic dipole moment with strength $\gamma_i$ shown as an explicit factor. $\rho$ is the density of the material. $e^s_{ab}(k)=\frac{1}{2}(\hat k_a\hat e^s_b+\hat k_b\hat e^s_a)$, $\hat k$ is the unit wavevector and $\hat e$ is the polarization unit vector. $r_i$ denotes the position of the $i^{\text th}$ TLS, and $\sigma_i^{x,z}$ are Pauli matrices. For simplicity we ignore the distinction between transverse and longitudinal TLS-phonon couplings.

Our model differs from the standard one in the distributions of the parameters $\varepsilon_i$, $\Delta^o_i$ and $\gamma_i$. In the standard model, the energy asymmetry between the right and left wells,  $\varepsilon_i$, and the tunneling matrix element $\Delta^o_i$ are assumed to  vary from site to site such that the probability per unit volume to find  a TLS with a given value of $\varepsilon_i$ and $\Delta_i^o$ is: 
\begin{equation}
  P(\varepsilon, \Delta^o)={\bar P}/\Delta^o
\end{equation}
with $0<\varepsilon<\varepsilon_{max}$, and  $\Delta^o_{min}<\Delta^o<\Delta^o_{max}$.  ${\bar P}= n_o/(\varepsilon_{max}\ln(\Delta^o_{max}/\Delta^o_{min}))$,  and $n_o$ is the density of TLS per unit volume. Typically, ${\bar P}$ is  an adjustable parameter fitted to experiments. The distribution of tunneling parameters is assumed to arise from a flat distribution of the tunneling  barrier heights \cite{Anderson_1972}. The coupling to phonons is chosen to be identical for all TLS, i.e., $\gamma_i=\gamma$, and is used as a fitting parameter.  The final assumption is that the interaction term between TLS and phonons is small, permitting the use of perturbation theory to compute the quantities above \cite{Phillips_1987}. 

In contrast to the standard model, we note that the phonon-TLS interaction implies that TLS can interact with one another via the strain field \cite{Yu_1988,Kassner_1989}. To effect this, we integrate out the phonons with energies higher than the tunneling matrix elements, i.e., $\hbar vk_m>\Delta_o^{max} \sim 10$ K.  (The precise value of $\Delta_o^{max}$ is not critical since it  ultimately only enters logarithmically.) The result of the integration is \cite{[{See Supplemental Material at [URL] }][{ for more details.}]supp_2019}:
\begin{equation}\label{eq:pol_H}
  \begin{split}
    H &= H_{\text{ph},k<k_m}+
{1\over 2}\sum_i\varepsilon_i\sigma_i^z+
{1\over 2}\sum_i\Delta^o_{max}e^{-\gamma_i^2/\gamma_o^2}\sigma_i^x \\
    &+ {1\over\sqrt{V}}\sum_{k<k_m,i}\gamma_i\epsilon^i_{k}\sigma^z_i
    +{1\over 2}\sum_{i\not= j}\sigma_i^zJ_{ij}\sigma_j^z
 \end{split}
\end{equation}
where
\begin{equation}\label{eq:gamma_o}
  \gamma_o={\sqrt{2}\over 3}\sqrt{\rho v^2v_o\hbar\omega_D}
\end{equation}
and $v_o$ is a unit volume of the chemical formula unit of the glass as would be used to define a  Debye frequency, $\omega_D$ \cite{Freeman_1986}. The third term shows that the tunneling matrix element has been renormalized downward by a polaron effect \cite{Kassner_1989} in which the overlap of the phonon wavefunctions between potential wells exponentially reduces the effective tunneling. Unlike the standard model where the tunneling depends on the WKB exponent that incorporates the barrier height, in our model the exponent of the tunneling matrix element depends on the TLS-phonon coupling $\gamma_i$. The fourth term contains the remaining TLS-phonon interaction which is weak and can be treated with perturbation theory as in the standard model. The last term shows that a TLS-TLS interaction term has been generated which is quite complex due to the tensorial nature of elastic dipole moments.  Following YL, $J_{ij}$ is simplified to  $J_{ij}={1\over\rho v^2}s_i\gamma_i s_j\gamma_j/r_{ij}^3$ where $r_{ij}$  is the distance between TLS $i$ and $j$, and $s_i=\pm 1$ is a spin representation of the orientation of the elastic dipoles (see \cite{Joffrin_1975} for the full expressions). 

To motivate our second assumption, let us review why the YL scenario failed to give the correct value of $C$. YL assumed $\gamma_i=\gamma\ \forall i$  so that $J_{ij}={\gamma^2\over\rho v^2}s_is_j/r_{ij}^3$ in Eq. (\ref{eq:pol_H}). The $1/r^3$ interactions together with a simple random mean field theory produces a density of states independent of the original density of TLS  given by:
\begin{equation}
  {\bar P}\approx {\rho v^2\over\gamma^2}
\label{eq:PbarSM}
\end{equation}
If we plug this into the expression for $C$, Eq. (\ref{eq:C_TLS}), we  get $C\sim 1$; universal but 4 orders of magnitude too large.  In addition, ${\bar P}$ is two orders of magnitude too large  compared to the density of states from specific heat measurements.  ${\bar P}$ could be reduced by increasing the strength of the  interactions $\gamma$ but this will not solve the $C\sim 1$ problem if the same value of $\gamma$ controls the attenuation of low frequency phonons. 

To fix this problem, we note that contrary to the standard model, the coupling between TLS and phonons should actually have a broad distribution. To see why, note that in Eq. (\ref{eq:H_TLS}), the TLS elastic dipole moment couples to the phonons via a $\sigma^z$ term, so that it is the {\it difference} between the dipole moments in the right and left wells of the TLS that couples to the strain field.  Assume now that the dipole moment in each well has the same magnitude ($\gamma_{max}$), but a different orientation. The difference between the dipole moments in the two wells is itself a dipole moment with magnitude $\gamma$. For two randomly oriented vector dipoles, the magnitude of the difference vector will have a uniform distribution $P(\gamma^2)=1/\gamma_{max}^2$ \cite{supp_2019}. The case of elastic tensor dipoles is more difficult but leads to a  similar distribution, albeit with an increase in probability for large values of $\gamma$ (see \cite{supp_2019}). 

The maximum possible value of the coupling, $\gamma_{max}$, is taken to be larger than the $\gamma$ inferred from acoustic experiments,  leading to stronger interactions between TLS and hence, a lower density of  states ${\bar P}$. In addition, these random TLS-phonons couplings produce a wide distribution of tunneling amplitudes due to the factor, $\exp{(-\gamma_i^2/\gamma_o^2)}$, multiplying $\Delta_o^{max}$.  TLS with large values of $\gamma_i$ have very small tunneling rates so they will not be seen, e.g., in internal friction measurements.  (The choice of a single value $\Delta_o^{max}$ for the tunneling matrix element prefactor is based on the view that while glasses have no obvious order,  any region is very similar to any other \cite{Malinovskii_1999}.) 

The next step is to approximate the TLS-TLS interaction term in  Eq. (\ref{eq:pol_H}) using a poor man's random mean field theory \cite{supp_2019}. The effective field felt by a given  TLS is the sum of the fields from all the surrounding TLS,  most of which are 'frozen' at low temperature:  $\varepsilon_i\equiv \varepsilon_i(J_{ij}=0)+\sum_{j\not= i}J_{ij}\sigma_j^z$.  Since we assume that the local asymmetry variations are small compared to  the interactions between TLS, we can neglect the $\varepsilon_i(J_{ij}=0)$  term so that the asymmetry energy $\varepsilon_i$ arises entirely from interactions. If we assume the $\sigma_j^z$ to be uncorrelated,  the sum will have approximately a Gaussian distribution with zero mean  (the $J_{ij}$ are equally positive and negative). The variance is given by: $Var(\varepsilon_i)\approx ({\gamma_{max}^2\over 2\rho v^2})^2 ({4\pi\over 3V_o})^2$ where $V_o$ is the average volume per rearranging  region. For energies small compared to the variance, the Gaussian  distribution is essentially flat and thus the probability of finding a  TLS with a given (small) $\varepsilon_i$ is  $P(\varepsilon_i)\approx 3\rho v^2V_o/(2\pi)^{2/3}\gamma_{max}^2$.  The density of states per unit energy and unit volume is then simply \cite{supp_2019}:
\begin{equation}\label{eq:dos1}
  {\tilde n}_o\approx \rho v^2/3\gamma_{max}^2
\end{equation}

With this random mean field approximation, the Hamiltonian in  Eq. (\ref{eq:pol_H}) reduces to that of an independent TLS model.  The effect of interactions between TLS has been subsumed into $n_o$, the distribution for the energy asymmetry per unit volume which is now  expressed in terms of material parameters. Together with the distribution of $\gamma$, $P(\gamma^2)=1/\gamma_{max}^2$, and the expression for  $\Delta^o$, $\Delta^o\equiv \Delta^o_{max}e^{-\gamma^2/\gamma_{max}^2}$,  we have an independent TLS model quite similar to the standard model.   The key difference is that $\gamma$ controls the value of the tunneling matrix element $\Delta^o$  in addition to the coupling between TLS and phonons.  It is more  convenient to change variables from ($\varepsilon$ and $\gamma$) to ($\varepsilon$ and $\Delta^o$). This gives: 
\begin{equation}\label{eq:dist}
  P(\varepsilon ,\Delta^o)= {{\bar P}\over\Delta^o}
\end{equation}
where
\begin{equation}\label{eq:Pbar}
  {\bar P}={1\over 3}{\rho v^2\over\gamma_{max}^2}
  ({\gamma_o\over\gamma_{max}})^2
\end{equation}
$\gamma$ is now given in terms of $\Delta^o$ by: 
\begin{equation}\label{eq:gamma}
  \gamma=\gamma_o\ln^{1/2}\left({\Delta^o_{max}\over\Delta^o}\right)={\sqrt{2}\over 3}\sqrt{\rho v^2v_o\hbar\omega_D}\ln^{1/2}\left({\Delta^o_{max}\over\Delta^o}\right)
\end{equation}

Let us bring all the pieces together and write our effective non-interacting Hamiltonian:
\begin{equation}\label{eq:final_H}
    H =H_{\text{phonon}}+ {1\over 2}\varepsilon\sigma^z +{1\over 2}\Delta^o\sigma^x 
    + \sum_{k<k_m}\gamma(\Delta^o)\epsilon_{k}\sigma^z
  \end{equation}
where $\gamma$ is an explicit function of $\Delta^o$ given by  Eq. (\ref{eq:gamma}) and the distribution of parameters is given by  Eq. (\ref{eq:dist}). Experimental quantities of interest should be computed with these expressions, though it is easier to do so by simplifying Eq. (\ref{eq:gamma}) for $\gamma$ as follows. With $\Delta^o_{max} \sim 10$ K, $\ln^{1/2}({\Delta^o_{max}\over\Delta^o})$ in $\gamma$ varies  from about 1.5 for $\Delta^o=1K$ to about 5 for $\Delta^o=5\times 10^{-11}K$  (which corresponds to an oscillation time of 1 second). Since $\Delta^o$  dictates which TLS can respond on an experimental timescales, $\gamma$  can be replaced by $\gamma_{eff}= \alpha\gamma_o$ with $\alpha$ equal to some constant in the range from 1 to 5. For concreteness, we will use $\alpha=2.5$.

With this simplification, we can now use Eq. (\ref{eq:C_TLS}) to calculate $C$ with $\gamma=\gamma_{eff}$ and Eq. (\ref{eq:Pbar}) for ${\bar P}$ to obtain:
\begin{equation}\label{eq:C_th}
    C={\bar P}{\gamma_{eff}^2\over\rho v^2} = 
{\alpha^2\over 3}({\gamma_o\over\gamma_{max}})^4
    \end{equation}
The last step is to estimate $\gamma_{max}$ which  requires going beyond elasticity. On general grounds, we  expect $\gamma_{max}\sim a\rho v^2v_o$ with $a<1$ being a material independent  constant \footnote{Numerical computations of elastic dipoles for  interstitials in crystals typically produce values for $\gamma$ of  order 10 - 20 eV \cite{Clouet_2018}.}. Using elastic stability criteria in disordered systems yields a better estimate \cite{Alexander_1998}:
    \begin{equation}\label{eq:gamma_max}
      \gamma_{max}= 
           {4\over 3}({2\over 9\pi})^{2/3}\rho v^2v_o \sim 0.23\rho v^2v_o
      \end{equation}
This and other ways to estimate $\gamma_{max}$ are further discussed in the Supplemental Material \cite{supp_2019}.

Table \ref{tbl:C_th_exp} shows the values of $C$ obtained from  Eq. (\ref{eq:C_th}) using Eq. (\ref{eq:gamma_max}) for $\gamma_{max}$ for the insulating glasses for which we have all the required data. We have  used $k_bT_D=\hbar\omega_D=\hbar v({6\pi^2\over v_o})^{1/3}$. $v_o$  is obtained from the material's chemical formula (see ref. 16 in  \cite{Berret_1988}) using $v_o=M/N_A\rho$ where M is the molecular mass  and $N_A$ is Avogardo's number. The only independent parameters are  $\rho$, $v_l$, $v_t$ and $v_o$. The theory has no adjustable parameters.

  \begin{table}
    \begin{tabular}{ c | c | c | c | c | c | c }
      \hline                  
      Glass & $\rho\ [kg/m^3]$ & $v\ [m/s]$ &$v_o\ [\AA^3]$ & $T_D\ [K]$ & $C_{exp}\cdot 10^4$ & $C_{th}\cdot 10^4$\\ \hline
      SiO$_2$ & 2200 & 4163 & 45.3 & 348 &  2.9 & 2.9 \\
      BK7 & 2510 & 4195 & 41.8 & 360 & 3.3 & 2.5 \\
      SF4 & 4780 & 2481 & 40.7 & 215 & 2.75 & 0.9 \\
      SF57 & 5510 & 2327  & 55.2 & 182 & 2.98 & 0.9 \\
      SF59 & 6260 & 2131  & 40.2 & 185 & 2.78 & 1.0 \\
      V52 & 4800 & 2511 & 61.1 & 190 & 4.9 & 0.8 \\
      BALNA & 4280 & 2569 &39.9 & 224 & 4.8 & 1.2 \\
      LAT & 5250 & 3105 & 68.2 & 226 & 3.7 & 0.3 \\
      Zn-glass & 4240 & 2580 & 45.9 & 215 & 3.6 & 2.0 \\
      PMMA & 1180 & 1762 & 138.4 & 101 & 3.7 & 2.9 \\
      \hline  
    \end{tabular}
    \caption{\label{tbl:C_th_exp} C for dielectric glasses computed 
from Eq. (\ref{eq:C_th}). Data from \cite{Berret_1988}.}
  \end{table}

While the overall comparison between theory and experiment are good, the discrepancies call for a discussion. First, we did not distinguish between longitudinal and transverse modes.  Given that experimentally \cite{Berret_1988} the ratio  $\gamma^2_l/v^2_l\approx \gamma_t^2/v^2_t$ and that it is the ratio that  matters for the TLS-phonon interaction, the errors from this approximation should not be large. In particular, this approximation cannot explain the large discrepancy for LAT between $C_{th}$ and $C_{exp}$  because $C_{th}$ for LAT is 10 times lower than $C_{th}$=$C_{exp}$ for SiO$_2$, even though the experimental difference between $\gamma_{l,t}$  and $v_{l,t}$ for the two materials is not large. A more likely source of the discrepancies is our estimate of the volume $v_o$ of the molecular formula unit which enters into the Debye temperature and is not well defined.  One possibility is to consider $v_o$ as the one adjustable  parameter of the theory.   

In short, the broad spectrum of TLS-phonon couplings $\gamma$ produces a distribution of tunneling parameters $\Delta^o$, many with values too small to contribute to ultrasonic measurements due to the exponential dependence of the tunneling on $\gamma^2$. The TLS that have tunneling amplitudes large enough to participate in ultrasonic experiments result in estimates of $\gamma$ of order 1 eV. This observed energy scale for $\gamma$ is consistent with $\gamma\sim\sqrt{\rho v^2v_o\hbar\omega_D}$ from Eq. (\ref{eq:gamma}).  For example, using values appropriate for SiO$_2$ ($\rho=2200$ kg/$m^3$, $v$=4200 m/s, $v_o=45\times 10^{-30}$ $m^3$ and $\hbar\omega_D=350K$), we find $\gamma\sim 0.57$ eV, in close agreement with the experimental values of $\gamma$ between 0.65 and 1 eV \cite{Berret_1988}. On the other hand, ${\bar P}$ is determined by the interaction between TLS, regardless of the amount of tunneling suppression. Eq. (\ref{eq:Pbar}) shows that the scale of ${\bar P}$ is dictated by $\rho v^2 v_o$ and $\gamma_{max}$, and hence is lower than what is found using Eq. (\ref{eq:PbarSM}) with the ultrasonic value $\gamma_{eff}$. This is why $Q^{-1}\sim 10^{-4}$ is so much smaller than in the orginal  YL approach. It should be possible to experimentally probe the distribution of $\gamma$ for TLS that couple to superconducting qubits and are altered by strain \cite{Grabovskij_2012}. 

Finally, since our effective Hamiltonian in Eq. (\ref{eq:final_H}) and the form of $P(\varepsilon,\Delta^o)$ in Eq. (\ref{eq:dist}) reduce to those of the standard TLS model, all the results of the standard model carry over with, at most, logarithmic temperature corrections of those quantities that depend on the TLS-phonon coupling since $\gamma$ has a logarithmic dependence on $\Delta^o$ as shown in Eq. (\ref{eq:gamma}).  In particular, the specific heat has the same temperature dependence as in the standard TLS model. The thermal conductivity $\kappa$ at low temperatures is limited by the scattering of phonons from TLS resulting in a logarithmic temperature correction: $\kappa\sim T^2/(1+2\ln(\Delta_{max}^o/2k_{B}T))$. 

Everything discussed so far applies for temperatures below 1K. Let us  briefly discuss what happens above $k_bT=\Delta^o_{max} \sim 10$ K.  The following estimate shows that the tunneling barrier height $V$ is comparable to $\Delta^o_{max}$. If we ignore the effect of phonons on tunneling, the bare tunneling matrix element is given in the WKB approximation by:
\begin{equation}
\Delta^o_{max}=\hbar\omega_De^{-\sqrt{2MV}d/\hbar}
\end{equation}
Solving for $V$ with  $\hbar\omega_D\sim 350K$ yields $\sqrt{2\rho v_oV}d/\hbar = \ln(35)$.  Using the numbers for SiO$_2$ with the barrier height $V$ in Kelvin and $d$ in $\AA$, we get  $1.6\sqrt{V}d\sim\ln(35)$, which means that $V\sim 5K$ for $d=1\AA$  and $V\sim 20K$ for $d=0.5\AA$. Thus, it is plausible that the barrier  height is in the 1-30K range which corresponds roughly to the  temperature where there is the plateau in the thermal conductivity and the boson peak in the specific heat. 

At temperatures much greater than the barrier height, thermal fluctuations make tunneling and the tunnel barrier irrelevant. So tunneling no longer reduces the density of states and thus, for $k_bT>>V$, we have ${\bar P}={\tilde n}_o={\rho v^2\over 3\gamma_{max}^2}$.  $\Delta^o$ decouples from $\gamma$ and the relevant coupling to phonons is  the average of $\gamma^2$ which is $\gamma_{max}^2/2$. Therefore, in this  regime, $C={\bar P}\langle\gamma^2\rangle/\rho v^2 = 1/6$ and is universal. The ratio of the mean free path to the wavelength becomes:
\begin{equation}
{\ell\over\lambda}={1\over 2\pi^2C}\sim 0.3
\end{equation}
This is observed in the thermal conductivity in the temperature range  above the plateau \cite{Freeman_1986}.  The intermediate temperature regime ($\sim 3-10$ K) corresponding to the plateau is very much material dependent and other processes come into play here \cite{Yu_1987}.  
  
In conclusion, we have elucidated aspects implicit in the standard TLS model that include strongly interacting TLS \cite{Yu_1988, Leggett_1991,Kassner_1989}, exponentially renormalized tunneling matrix elements \cite{Kassner_1989}, and a heretofore unrecognized broad distribution of TLS-phonon couplings. This produces the correct order of magnitude for $Q^{-1}$ and the coupling $\gamma$ seen in acoustic experiments. Variations in the predicted values of $Q^{-1}$  from material to material are only slightly larger than in experiments.  At high temperatures, where tunneling is irrelevant,  we predict $\ell/\lambda \sim 1$, consistent with thermal conductivity experiments. 

\begin{acknowledgments}
This research was primarily supported by the National Science Foundation (NSF) through the University of Wisconsin Materials Research Science and Engineering Center (DMR-1720415). The work of CCY was performed in part at the Aspen Center for Physics, which is supported by NSF grant PHY-1607611.
\end{acknowledgments}


%
\end{document}


\title{Supplementary Material for\\ ``Why phonon scattering in glasses is 
universally small at low temperatures''}
\author{Herve Carruzzo}
\author{Clare C. Yu}
\affiliation{University of California, Irvine}
\date{\today}

\maketitle

\section{Distribution of $\gamma$}\label{app:gamdist}
The coupling between phonons and TLS is of the form
\begin{equation}\label{eq:H_int}
  H_{int}=\gamma D\sigma_z\epsilon
\end{equation}
where $\gamma D\sigma_z$ is the elastic dipole operator. This operator is a projection onto the right and left basis states $|\psi_{R,L}\rangle$ of the full dipole operator $\boldsymbol{D}$. 
\begin{equation}
  \begin{split}
  \boldsymbol{D}=&
  \begin{bmatrix} D^{LL} & D^{LR}\\ D^{RL} & D^{RR}\end{bmatrix}\\
  =& {1\over 2}(D^{LL}+D^{RR})\boldsymbol{I}+{1\over 2}(D^{LL}-D^{RR})\sigma^z+D^{LR}\sigma^x
  \end{split}
  \end{equation}

  \noindent were $D^{AB}\equiv\langle \psi_A|\boldsymbol{D}|\psi_B\rangle$. $I$ is the identity matrix and thus can be dropped. $D^{LR}=D^{RL}$ depends on the overlap between the right and left wells and hence, is taken to be negligible. As a result, only the $\sigma^z$ term remains, which is what goes into Eq. (\ref{eq:H_int}). {\it The key is that the $\sigma^z$ term is half the difference of the elastic dipole moment $D$ in the right and left wells.}

Assuming the magnitude $\gamma_{max}$ of the elastic dipole to be the same in the right and left well, what can be said about the distribution of the (half) difference $\Delta D_{\alpha\beta}$, assuming some distribution for the relative ``orientation'' of the dipoles?

The case of vector dipoles, which is directly relevant for dielectric experiments, is instructive. 
 Choose the dipole in the right well, say, to be along the z axis and the dipole on the left well to be randomly oriented in the $\hat n$ direction on the unit sphere. The difference of two unit vectors is a vector that need not have unit length. The length of the new vector will multiply $\gamma_{max}$ and dictate the magnitude of the dipole difference $\gamma$ between the two wells. The probability distribution for $P(\gamma^2=\gamma_{max}^2|\hat z - \hat n|^2/4)$ is:

  \begin{equation}
    \begin{split}
      P(\gamma^2)&={1\over 4\pi}\int_0^{2\pi}d\phi\int_0^{\pi}d\theta\sin(\theta)\delta(\gamma^2-\gamma^2_{max}(1-\cos(\theta))/2)\\
      &= {1\over \gamma_{max}^2}\ \ \ 0<\gamma<\gamma_{max}
      \end{split}
\label{eq:vectorDipole}
    \end{equation}
or equivalently $P(\gamma)=2\gamma/\gamma_{max}^2$. 

Now consider the elastic dipole. 
While the most general case is $D_{ij}=an_in_j+bm_im_j$, with $\vec n$ and 
$\vec m$ being two unit vectors and $a$ and $b$ being constants, it is easier consider the simpler case of $D_{ij}=\gamma_{max}n_in_j$ where $\vec n$ is a unit vector. Again we pick $\vec n$ to be along the z axis in well 1 and randomly distributed on the unit sphere in well 2. Then $D^{(1)}_{ij}=\gamma_{max}\delta_{zi}\delta_{zj}$ and $D^{(2)}=R^{-1}D^{(1)}R$ where $R$ represents an arbitrary rotation. Thus $\Delta D=D^{(1)}-R^{-1}D^{(1)}R$. The first operation of R is a rotation by an angle $\theta$ around a vector in the xy plane. The second operation is a rotation around the z axis which leaves the shape of $\Delta D$ unchanged so that the shape of $\Delta D$ only depends on $\theta$. For the particular triad $\hat x,\; \hat y,\; \hat z$ that diagonalizes $\Delta D$, we get:
\begin{equation}
  \Delta D=\gamma_{max}\sqrt{1-\cos^2(\theta)}
  \begin{bmatrix} 0 & 0 & 0\\ 0 & -1 & 0\\
                  0 & 0 & 1
\end{bmatrix}
  \end{equation}
  This implies that $\Delta D$ is of the form $\gamma_{max}\sin(\theta)(\hat z_i\hat z_j-\hat y_i\hat y_j)$, which is more general than the original tensor but of the same norm. On that basis, what matters is the distribution of the prefactor.  The probability distribution of half the magnitude of $\Delta D$ is then:
  \begin{equation}
    \begin{split}
    P(\gamma)=&{1\over 2}\int_0^{\pi}d\theta\sin(\theta)\delta(\gamma - \gamma_{max}\sin(\theta)/2)\\
       =& {4\gamma\over\gamma_{max}^2\sqrt{1-4\gamma^2/\gamma_{max}^2}},\ \ 0<\gamma< \gamma_{max}/2
       \end{split}
    \end{equation}
Given that the deviation from linearity only occurs for large values of $\gamma$, the linear approximation $P(\gamma)\approx 4\gamma/\gamma^2_{max}$
is reasonable. Note that this expression and the normalizing value of 
$\gamma_{max}$ differs from that found in Eq. (\ref{eq:vectorDipole}) 
because we are dealing with elastic stress dipoles here rather than vector dipoles. 

\section{Integrating out the phonons\label{app:DOT}}

In a system of TLS interacting with phonons, the timescale describing the TLS is much slower than that of phonons with energies $\hbar\omega(k)\gg \Delta^o$ where $\omega(k)$ is the phonon spectrum and $\Delta^o$ is the tunneling matrix element of the TLS. Since the tunneling matrix element is at most a few Kelvin, the vast majority of the phonons are fast compared to the TLS. This situation allows us to integrate out the fast phonons from the problem, resulting in an effective low energy Hamiltonian with the TLS parameters renormalized. We detail here the steps involved. The starting point is the Hamiltonian for an ensemble of TLS interacting with phonons:

\begin{equation}\label{eq:H_TLS}
    H=H_{\text{ph}}+{1\over 2}\sum_i 
(\varepsilon_i\sigma^z_i + \Delta_i^o\sigma_i^x) + 
{1\over\sqrt{V}}\sum_{ik}\gamma_i \epsilon^i_{k}\sigma^z_i
\end{equation}
where the free phonon Hamiltonian  $H_{\text{ph}}=\sum_{ks}\hbar\omega(ks)\psi^{\dagger}_{ks}\psi_{ks}$.  $\psi^{\dagger}_{ks}$ and $\psi_{ks}$ are the phonon creation and anihilation  operators for mode $k$ and polarization $s$. In the TLS-phonon interaction (last term in Eq. (\ref{eq:H_TLS})), $\epsilon^i_{k} = \sum_s\xi_{ks}(i)\psi_{ks}+\xi_{ks}^*(i)\psi_{ks}^{\dagger}$ represents the scalar strain field, where $\xi_{ks}=i\sqrt{ {\hbar\omega(ks)\over 2\rho v^2} } D^{i}_{ab}e^s_{ab}(k)e^{ikr_i}$ and $\gamma_i D_{ab}^{i}$ is the TLS elastic dipole moment with strength $\gamma_i$ shown as an explicit factor.  $\rho$ is the density of the material and $v$ is the speed of sound. $e^s_{ab}={1\over 2}(\hat k_a\hat e^s_b+\hat k_b\hat e^s_a)$, $\hat k$ is the unit wavevector and $\hat e$ is the polarization unit vector. $r_i$ denotes the position of the $i^{\text th}$ TLS, and $\sigma_i^{x,z}$ are Pauli matrices. $\varepsilon_i$ and $\Delta^o_i$ are the asymmetry energy and tunneling matrix element of the $i^{\text th}$ TLS respectively. 

Since the Hamiltonian Eq. (\ref{eq:H_TLS}) is at most quadratic in the phonon fields, integrating out the high frequencies amounts to a shift of the phonon coordinate (by completing the square). The trick is that this shift needs to be different for each state (i.e., the right and left well states), since the linear coupling has the opposite sign (because of the $\sigma_i^z$ operator). This shifting of the coordinates is achieved by a unitary transformation (also known as a polaron transformation, see, e.g., \cite{Kassner_1989}):
\begin{equation}\label{eq:polaron} U=e^{-{1\over\sqrt{V}}\sum_{ks,i}\sigma^z_i\gamma_i(\xi_{ks}(i)\psi_{ks}-\xi^*_{ks}(i)\psi^{\dagger}_{ks})/(\hbar\omega(ks))}
\end{equation}
The operators in the old and new basis are related by
\begin{eqnarray}
 \phi_{ks}&\equiv &U^{-1}\psi_{ks}U \nonumber \\ 
 S^n &\equiv& U^{-1}\sigma^n U
\label{eq:phiS}
\end{eqnarray}
where $n=x,z$. 
Explicitly:
\begin{equation}\label{eq:psi}
  \psi_{ks}=\phi_{ks}-\sum_iS_i^z{\gamma_i\xi^*_{ks}(i)\over\hbar\omega(ks)}
\end{equation}
where we have used the relation $[f(A),B]=[A,B]{\partial f\over\partial A}$, the fact that $\sigma_z$ commutes with $U$, and:
  \begin{equation}\label{eq:newsigma}
       \sigma_i^x=\cosh(2\alpha)S_i^x 
       +\sinh(2\alpha)S_i^zS_i^x     
   \end{equation}
where $\alpha={1\over\sqrt{V}}\sum_{ks}\gamma_i(\xi_{ks}(i)\psi_{ks}-\xi^*_{ks}(i) \psi_{ks}^{\dagger})/(\hbar\omega(ks))$.
Using:
\begin{equation}
     e^{A+B}=e^Ae^Be^{-{1\over 2}[A,B]}
   \end{equation}
which is valid if $[A,[A,B]]=[B,[A,B]]=0$, we can normal order the Bose operators in Eq.(\ref{eq:newsigma}):
\begin{equation}
  \begin{split}
  :\cosh(2\alpha):\ =& e^{-\sum_{ks}2\tilde\xi^*_{ks}\tilde\xi_{ks}}(e^{-\sum_{ks} 2\tilde\xi^*_{ks}\phi^{\dagger}_{ks}}e^{\sum_{ks}2\xi_{ks}\phi_{ks}} \\ &+e^{\sum_{ks}2\tilde\xi^*_{ks}\phi^{\dagger}_{ks}}e^{-\sum_{ks}2\tilde\xi_{ks}\phi_{ks}})
  \end{split}
\end{equation}
Likewise:
\begin{equation}
  \begin{split}
  :\sinh(2\alpha):\ =& e^{-\sum_{ks}2\tilde\xi^*_{ks}\tilde\xi_{ks}}(e^{-\sum_{ks} 2\tilde\xi^*_{ks}\phi^{\dagger}_{ks}}
  e^{2\tilde\xi_{ks}\phi_{ks}} \\
  &-e^{\sum_{ks} 2\tilde\xi^*_{ks}\phi^{\dagger}_{ks}}e^{-\sum_{ks} 2\tilde\xi_{ks}\phi_{ks}})
  \end{split}
\end{equation}
where $\tilde\xi_{ks}=\gamma_i\xi_{ks}/\sqrt{V}\hbar\omega(ks)$. Thus $:\cosh(2\alpha):=e^{-\sum_{ks} 2\tilde\xi^*_{ks}\tilde\xi_{ks}}(1-4\sum_{ks}\tilde\xi^*_{ks}\tilde\xi_{ks}\phi_{ks}^{\dagger}\phi_{ks}+ 4\sum_{ks}\tilde\xi_{ks}^2\phi_{ks}^2+ 4\sum_{ks}(\tilde\xi_{ks}^*)^2(\phi_{ks}^{\dagger})^2+ O(\phi^3))$ and $:\sinh(2\alpha):=e^{-\sum_{ks} 2\tilde\xi^*_{ks}\tilde\xi_{ks}}(4\sum_{ks}\tilde\xi_{ks}\phi_{ks}- 4\sum_{ks}\tilde\xi_{ks}^*\phi_{ks}^{\dagger}+O(\phi^3))$. 

Since the phonons are much faster than TLS tunneling rate, we need only keep
zeroth order in the phonon fields. This gives:
\begin{equation}\label{eq:sx}
\sigma_i^x=e^{-{1\over V}\sum_{ks} 2\gamma_i^2\xi^*_{ks}\xi_{ks}/(\hbar\omega(ks))^2}S_i^x
\end{equation}  
Evaluating the sum in the exponent yields ${\gamma_i^2\over\gamma_o^2}(1-({vk_m\over\omega_D})^2)$ where $ \gamma_o={\sqrt{2}\over 3}\sqrt{\rho v^2v_o\hbar\omega_D}$. Since $vk_m$ is less than one tenth of $\omega_D$ in the present model, $k_m$ can be set to zero in this expression. Finally, substituting Eqs. (\ref{eq:phiS}) and Eq. (\ref{eq:sx}) into Eq. (\ref{eq:H_TLS}) as well as relabelling
$S^n_i$ by $\sigma^n_i$ and $\phi_k$ by $\psi_k$ gives the result:
\begin{equation}\label{eq:pol_H}
  \begin{split}
    H &= H_{\text{ph},k<k_m}+
{1\over 2}\sum_i\varepsilon_i\sigma_i^z+
{1\over 2}\sum_i\Delta^o_{max}e^{-\gamma_i^2/\gamma_o^2}\sigma_i^x \\
    &+ {1\over\sqrt{V}}\sum_{k<k_m,i}\gamma_i\epsilon^i_{k}\sigma^z_i
    +{1\over 2}\sum_{i\not= j}\sigma_i^zJ_{ij}\sigma_j^z
 \end{split}
\end{equation}
with $J_{ij}$ given by:
\begin{equation}
  J_{ij}={1\over V}\sum_k{\gamma_i\xi_k^*(i)\gamma_j\xi_k(j)\over\hbar\omega(k)}
  \end{equation}
\section{Poor man's mean field approach to estimating the density of states 
$n_o$ \label{app:RMFT}}
We review here the naive random mean field used for calculating the TLS density of states ${\tilde n}_o$. We note that this approach gives incorrect results for all quantities involving correlation functions. However, for the purpose of computing the local effective field for low energies of a TLS, it should be fine due to the large distribution of fields obtained. (It should also be noted that at very low energies, a gap in the density of states must develop for stability reasons \cite{Burin_1995}. This is analogous to the formation of the Coulomb gap that arises in electron glasses \cite{Pollak_1970,Efros_1975,Shklovskii_1984}. This gap occurs well below the energies and hence, temperatures considered in this work.) 

The starting point is a Hamiltonian of the form:
\begin{equation}
H={1\over 2}\sum_{i\not=j}J_{ij}\sigma_i^z\sigma_j^z
\end{equation}
where
\begin{equation}\label{eq:J_ij}
  J_{ij}={\gamma_ix_i\gamma_jx_j\over\rho v^2r_{ij}^3}
  \end{equation}
$\gamma_i$ are the magnitude of the couplings between phonons and TLS. These couplings are distributed between 0 and some maximum value $\gamma_{max}$ such that $P(\gamma^2)$ equals a constant. $x_i$ is meant as a simplified version of full TLS elastic dipole tensor with random orientation.  The distribution of $x_i$ is such that $<x_i>=0$ and $<x_i^2>=1$. The requirement $<x_i^2>=1$ comes from the fact that experiments only measure averages over the square of the elastic dipole moment's orientation. $r_{ij}$ is the distance between two TLS randomly distributed at positions $r_i$ and $r_j$. $\sigma^z_i$ is the $z$th Pauli matrix.

In the spirit of the Weiss mean field approach, the local field is written as:
  \begin{equation}
    h_i= \sum_jJ_{ij}m_j
  \end{equation}
where $m_i$ is the ``magnetization'' induced by the field $h_i$. $m_i=\tanh(\beta h_i)$ where $\beta$ is the inverse temperature $1/k_BT$. If the distribution of values for $h_i$ is large, then at low temperatures, most TLS will be saturated (with $m_j=\pm 1$), and those TLS that are not saturated will tend to be isolated and thus will not contribute to the sum. We assume that $h_i$ has a Gaussian distribution $p(h_i)={1\over\sqrt{2\pi\sigma^2}}e^{-h_i^2/2\sigma^2}$. All we need is the variance of this distribution:
  \begin{equation}
    \sigma^2 = \langle (h_i)^2 \rangle = \sum_{jk}\langle J_{ij}J_{ik}m_jm_k\rangle
  \end{equation}
  where $\langle ... \rangle$ denotes averages over the disorder. We next assume that the J's and m's are uncorrelated and use the fact that $m_i^2=1$ to get:
  \begin{equation}
     \sigma^2  = \sum_j \langle J_{ij}^2 \rangle
   \end{equation}
   Using the definition for $J_{ij}$ in Eq. (\ref{eq:J_ij}) yields:
   \begin{equation}
     \sigma^2 = \left({1\over\rho v^2}\right)^2\left\langle \sum_j{\langle x_i^2 \rangle\langle x_j^2 \rangle \langle \gamma_i^2 \rangle \langle \gamma_j^2 \rangle\over  r_{ij}^6}\right\rangle_{r_j} 
     \end{equation}
where $\langle...\rangle_{r_j}$ is an average over all possible positions of
the TLS. Now $\langle x_i^2 \rangle=1$ and $P(\gamma_i^2)=1/\gamma_{max}^2$ with $\gamma_i\in [0,\gamma_{max}]$, so $\langle \gamma_i^2 \rangle=\gamma_{max}^2/2$. Therefore:
     \begin{equation}
       \sigma^2 = \left({\gamma_{max}^2\over 2\rho v^2}\right)^2\left\langle\sum_j {1\over r_{ij}^6}\right\rangle_{r_j}
       \end{equation}
       Let us define $R_o$ such that ${4\over 3}\pi R_o^3$ is the volume $V_o$ per TLS and replace the average of the sum by an integral: $\langle\sum_i...\rangle_{r_i}\rightarrow {1\over V_o}\int_{R_o}^{\infty}4\pi r^2... dr$. Thus:
       \begin{equation}
         \left\langle\sum_j {1\over r_{ij}^6}\right\rangle_{r_j}\approx {1\over V_o}\int_{R_o}^{\infty}{4\pi r^2\over r^6}dr = \left({4\pi\over 3V_o}\right)^2
         \end{equation}
         The variance is then $\sigma^2=({\gamma_{max}^2\over 2\rho v^2})^2({4\pi\over 3V_o})^2$. For the small values of $h_i$ that we are interested in, we can set $h_i=0$ in the Gaussian distribution: $P(h_i)\approx 1/\sqrt{2\pi\sigma^2}\approx {\rho v^2\over\gamma_{max}^2}{3V_o\over (2\pi)^{3/2}}$. Finally, to obtain the density of states $n_o$, we have to divide by the volume per TLS, $V_o$. We also need to remember that in the TLS model, the energy asymmetry is taken to be positive, so we must 'fold' the negative values onto the positive ones, which gives a factor 2. The result is then:
         \begin{equation}
           {\tilde n}_o= {\rho v^2\over\gamma_{max}^2}{6\over (2\pi)^{3/2}}\approx  {\rho v^2\over 3\gamma_{max}^2}.
           \end{equation}
where we have approximated ${6\over (2\pi)^{3/2}}\sim 0.38...$ by 1/3.  

\section{Estimating $\gamma_{max}$ Using  Stability criterion}
\label{app:Stability}

In this section we present various ways to estimate $\gamma_{max}$.
The maximal value of a dipole moment is governed by what are essentially stability arguments. The elastic medium has to be able to handle the stress from the local dipole. It then boils down to figuring out what is the maximal atomic strain allowed. The first step is to estimate the strain produced at the atoms neighboring a given dipole of strength $\gamma$ at the origin. That is easy to do. The result is:
\begin{equation}
  \delta^2={3\gamma^2\over 2(\rho v^2v_o)^2}
  \end{equation}
$\delta$ represents a typical strain created in the surrounding medium by the dipole. The question has thus shifted to figuring out what the maximum $\delta$ should be. There are a few options: (In what follows, any numerical estimate is done for SiO$_2$.)
 \begin{table*}[t] 
    \begin{tabular}{ c | c | c | c | c | c | c | c | c | c | c | c }
      \hline                  
      Glass & ${\rho\atop [kg/m^3]}$ & ${v\atop [m/s]}$ &${v_o\atop [\AA^3]}$ & ${T_g\atop [k]}$ & ${T_D\atop [K]}$ & ${C_{exp}\atop 10^4}$ & ${{\bar P}_{exp}\atop [10^{45}/J\cdot m^3)]}$ & ${C_{g}\atop 10^4}$ & ${{\bar P}_{g}\atop [10^{45}/J\cdot m^3)]}$ & ${C_{e}\atop 10^4}$ & ${{\bar P}_{e}\atop [10^{45}/J\cdot m^3)]}$ \\ \hline
      SiO$_2$ & 2200 & 4163 & 45.3 & 1473 & 348 & 2.9 & 0.8 & 2.8 & 0.9 & 2.9 & 1.0\\
      BK7 & 2510 & 4195 & 41.8 & 836 & 360 & 3.3 & 1.1 & 9.2 & 3.1 & 2.5 & 0.9\\
       SF4 & 4780 & 2481 & 40.7 & 693 & 215 & 2.75 & 1.1 & 3.8 & 1.8 & 0.9 & 0.4 \\
       SF57 & 5510 & 2327  & 55.2 & 687 & 182 & 2.98 & 1.0 & 3.6& 1.8 & 0.9 & 0.5 \\
       SF59 & 6260 & 2131  & 40.2 & 635 & 185 & 2.78 & 1.0 & 3.7 & 2.2 & 1.0 & 0.6 \\
       V52 & 4800 & 2511 & 61.1 & 593 & 190 & 4.9 & 1.7 & 5.2 & 2.3 & 0.8 & 0.3\\
       BALNA & 4280 & 2569 &39.9 & 520 & 224 & 4.8 & 2.1 & 7.6 & 3.6 & 1.2 & 0.6\\
       LAT & 5250 & 3105 & 68.2 & 723 & 226 & 3.7 & 1.4 & 5.0 & 1.7 & 0.3 & 0.1\\
       Zn-glass & 4240 & 2580 & 45.9 & 570 & 215 & 3.6 & 2.2 & 7.2 & 3.8 & 2.0 & 1.1\\
       PMMA & 1180 & 1762 & 138.4 & 374 & 101 & 3.7 & 0.6 & 3.7 & 1.4 & 2.9 & 1.1\\
      \hline  
    \end{tabular}
    \caption{\label{tbl:g_vs_e} Predictions for $C$ and ${\bar P}$ for dielectric glasses using two ways of estimating $\gamma_{max}$ as described in the text. Data from \cite{Berret_1988} and references therein.}
  \end{table*}

  \begin{enumerate}
  \item  The elastic medium becomes unstable well before the average displacement of its constituents reaches the lattice spacing. So assume that the strain $\delta$ has some fixed value $\alpha_1 \sim 0.2-0.25$. Then $\gamma_{max}=\alpha_1\sqrt{{2\over 3}}\rho v^2v_o$. For $C$ to be $\sim 10^{-4}$, then $\alpha_1$ needs to be $1/4$. This is quite a reasonable number.
  \item A better estimate might be obtained from a ``reverse Lindemann criterion''. The maximum strain fluctuations that the glass can sustain occur at the glass transition temperature $T_g$. So $\delta$ should be set as a multiple of these maximal strain fluctuations which are given by: 
    \begin{equation}
      \langle e^2 \rangle_{T>T_D} = {3\hbar\omega_D\over\rho v^2 v_o}{T_g\over T_D}
    \end{equation}
where $e$ is the strain field. Taking $\langle e^2 \rangle_{T>T_D}$ as the maximum value for $\delta^2$ gives the following estimate for $\gamma_{max}$:
    \begin{equation}
      \gamma_{max}=\alpha_2\sqrt{2\rho v^2v_ok_bT_g}
    \end{equation}
    where $\alpha_2$ is a proportionality factor of order 1. This estimate for $\gamma_{max}$ gives $C\sim 10^{-4}$ with $\alpha_2=1$, but to fit $C$ to the value for SiO$_2$ requires $\alpha_2=1.5$. It is noteworthy that Lubchenko et al. \cite{Lubchenko_2008} have obtained the same expression (without the factor 2) for $\gamma_{max}$ based on a slightly different elastic stability argument. 
  \item The expression for $\langle e^2 \rangle$ was written on purpose in an odd way proportional to $\hbar\omega_D/T_D$. The reason is that the quantity that multiplies $T_g/T_D$ is just $8/3$ times the expression for $\langle e^2 \rangle_{T=0}$. It thus seems natural to think that an equally good approximation would be to approximate $\delta$ as a multiple of $\sqrt{\langle e^2 \rangle_{T=0}}$. In this case, the estimate for $\gamma_{max}$ becomes:
    \begin{equation}
      \gamma_{max}=\alpha_3\sqrt{\rho v^2v_o\hbar\omega_D}
    \end{equation}
In this case, $\gamma_{max}$ is directly proportional to $\gamma_o$ and $C$ becomes entirely material independent. $\alpha_3$ needs to be about 5 in order to get $C\sim 10^{-4}$.
\item  Another way to estimate $\gamma_{max}$ is to use Alexander's elasticity theory \cite{Alexander_1998} of disordered systems. In his model, internal stresses permeate the glass and, in particular, negative bond tension, i.e., compression, can occur that will lead to buckling. In this case, the energy of a bond of length R is ${T\over 2R}u^2+{k\over 4R^2}u^4$ where $u$ is the deviation from the normal length R, $T$ is the bond tension and k is the elastic spring constant which is $\omega_D\rho v_o$. When the tension $T$ is negative (corresponding to compression), the bond buckles, resulting in $u=\sqrt{{2TR\over k}}$. Using $\omega_D=vk_D$ and $k_D=({9\pi\over 2})^{1/3}/R$, the strain $e$ can be expressed as $e=({2\over 9\pi})^{1/3}\sqrt{{2TR\over\rho v^2v_o}}$. Equating $TR$ to $\gamma_{max}$, and $e$ to $\delta$, i.e., stating that the elastic medium should support the strain $e$, we obtain 
    \begin{equation}
      \gamma_{max}={4\over 3}({2\over 9\pi})^{2/3}\rho v^2v_o
    \end{equation}
    Again, we should assume that there is a factor of proportionality of order one. This is the expression used in the main text. 
\end{enumerate}
    Table \ref{tbl:g_vs_e} lists the predictions for ${\bar P}$ and $C$ using the glass transition estimate, ${\bar P}_g$ and $C_g$, with $\alpha_3 = 1.5$, as well as the prediction using the estimate from stability, ${\bar P}_e$ and $C_e$. In each case, both ${\bar P}$ and $C$ show fluctuations consistent with the experimental data. The values derived from the estimate based on the glass transition temperature are closer to the experimental values but also show more significant deviations in some cases, e.g., BK7 and Zn-glass. All these various estimates yield the correct order of magnitude for $C$ and ${\bar P}$ which underscores the robustness of our approach. 
    
%